\documentclass[12pt,twoside]{article}
\usepackage{fleqn,espcrc1}
\usepackage{graphicx}

\newcommand{\snn}{\sqrt{s_{NN}}}
\newcommand{\pp}{pp}
\newcommand{\pbarp}{\overline{p}p}
\newcommand{\pA}{pA}

\newcommand{\np}{N_{part}}
\newcommand{\nc}{N_{coll}}
\newcommand{\ns}{N_{spec}}
\newcommand{\half}{\frac{1}{2}}
\newcommand{\halfnp}{\half\np}

\newcommand{\etazero}{\eta = 0}
\newcommand{\etaone}{|\eta| < 1}
\newcommand{\nch}{N_{ch}}
\newcommand{\et}{E_T}
\newcommand{\pt}{p_T}
\newcommand{\vone}{v_1}
\newcommand{\vtwo}{v_2}

\newcommand{\dndeta}{d\nch/d\eta}
\newcommand{\detdeta}{d\et/d\eta}
\newcommand{\dndy}{d\nch/dy}
\newcommand{\dndetazero}{\dndeta|_{\etazero}}
\newcommand{\dndetaone}{\dndeta|_{\etaone}}
\newcommand{\dndetanp}{\dndeta / \halfnp}
\newcommand{\dndetazeronp}{\dndetazero / \halfnp}
\newcommand{\dndetaonenp}{\dndetaone / \halfnp}

\usepackage{graphicx}

\title{Global Observables at RHIC}

\author{Peter A. Steinberg
\address
	{ 
        Brookhaven National Laboratory\\ 
        Upton, NY 11973
	}	
}	
       
\begin{document}

\maketitle

\begin{abstract}
The first three measurements from the RHIC program were
results on global observables: charged particle multiplicity ($\nch$),
transverse energy ($\et$) and elliptic flow ($\vtwo$). 
They offer a look at the large-scale features of particle production in
high-energy nuclear collisions, with particular insight into entropy
production and collective behavior.
Results from all of the RHIC experiments are discussed in light of data from
lower energy nuclear collisions as well as from high-energy hadronic
collisions to test our current understanding of the collision dynamics.
\end{abstract}

\section{Introduction}

It is hoped that Au+Au collisions at RHIC will form a large, collective state
of thermally equilibrated matter.  
Such a state should be describable
by simple quantities, 
such as energy density and pressure, 
which can be related to lattice or hydrodynamic calculations 
incorporating a nuclear equation of state.
One means to study this is to 
measure global quantities that characterize the entire event.
Looking along the beam axis, we can attempt to
understand the energy density achieved in the collision by 
studying the multiplicity and $E_T$ distributions ($\dndeta$, $\detdeta$). 
These can also be compared with $\pp$, $\pbarp$ and $\pA$
collisions at higher and lower energies to study
their scaling with the number of participating nucleons as well
as the number of binary collisions.   
Azimuthal distributions in slices of $\eta$, 
studied by means of their Fourier
coefficients, can be compared with results from 
lower energy nuclear collisions,
where a steady increase of $\vtwo$ with energy has been observed.
This indicates that the particle emission becomes increasingly ellipsoidal,
suggesting increased transverse pressure.
Taken together, these measurements offer the first look at 
collective particle production and dynamics at RHIC.
It is an open question, however, whether features of the initial
state can survive the hadronization process.
We review the data discussed
at Quark Matter 2001 \cite{qm2001} to test our understanding of how the dynamics
of the early stages of RHIC collisions can be observed in the final state.

\section{Centrality} 

\begin{figure}[t]
\begin{minipage}[t]{75mm}
\includegraphics[width=8cm]{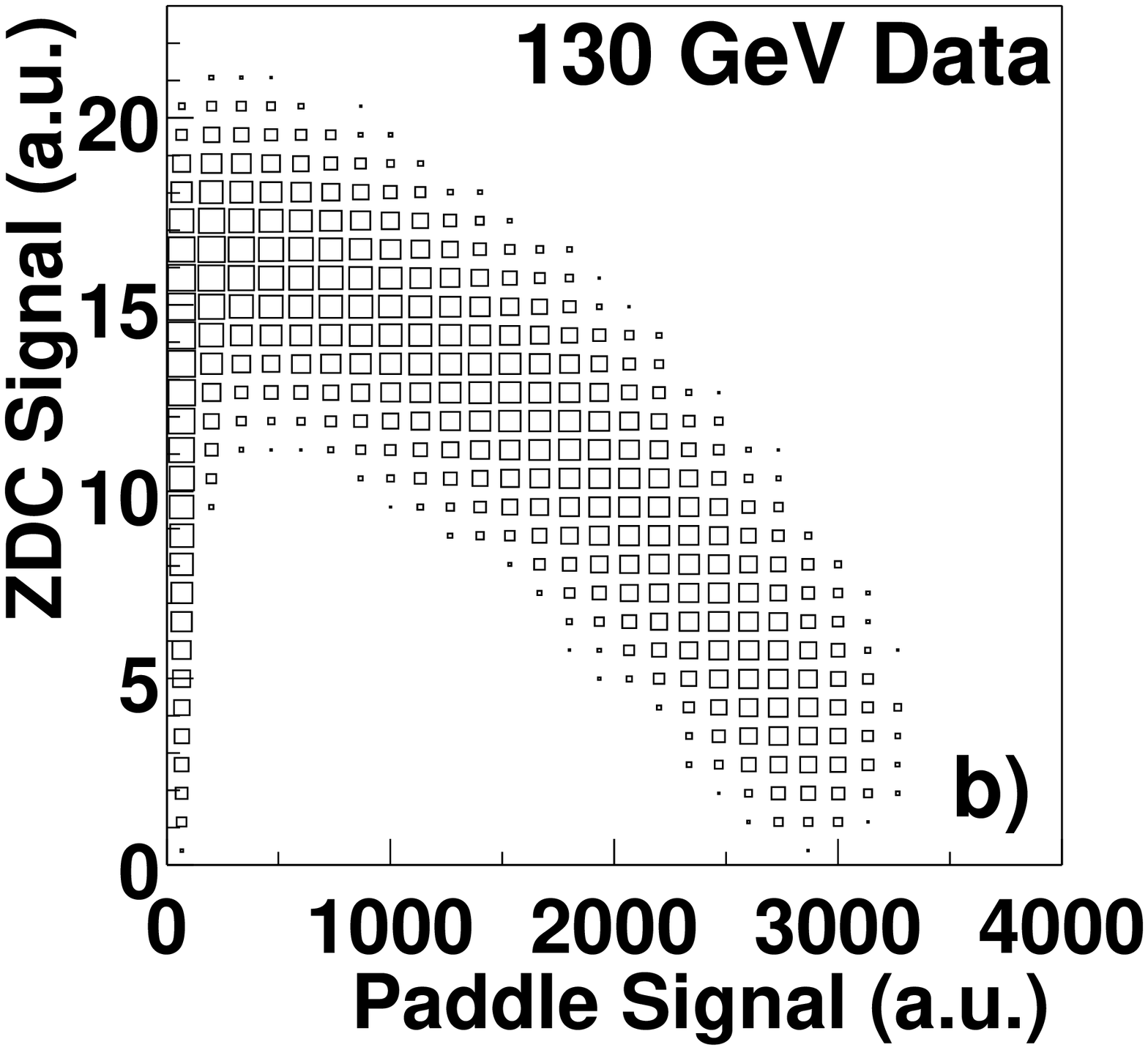}
\vspace*{-1cm}
\caption{\small Correlation between the ZDC energy and multiplicity 
	between $3<|\eta|<4.5$ (from PHOBOS).}
\label{zdc_vs_pdlmean}
\end{minipage}
\hspace{\fill}
\begin{minipage}[t]{75mm}
\raisebox{+.5cm}
{
\includegraphics[width=8cm,height=7cm]{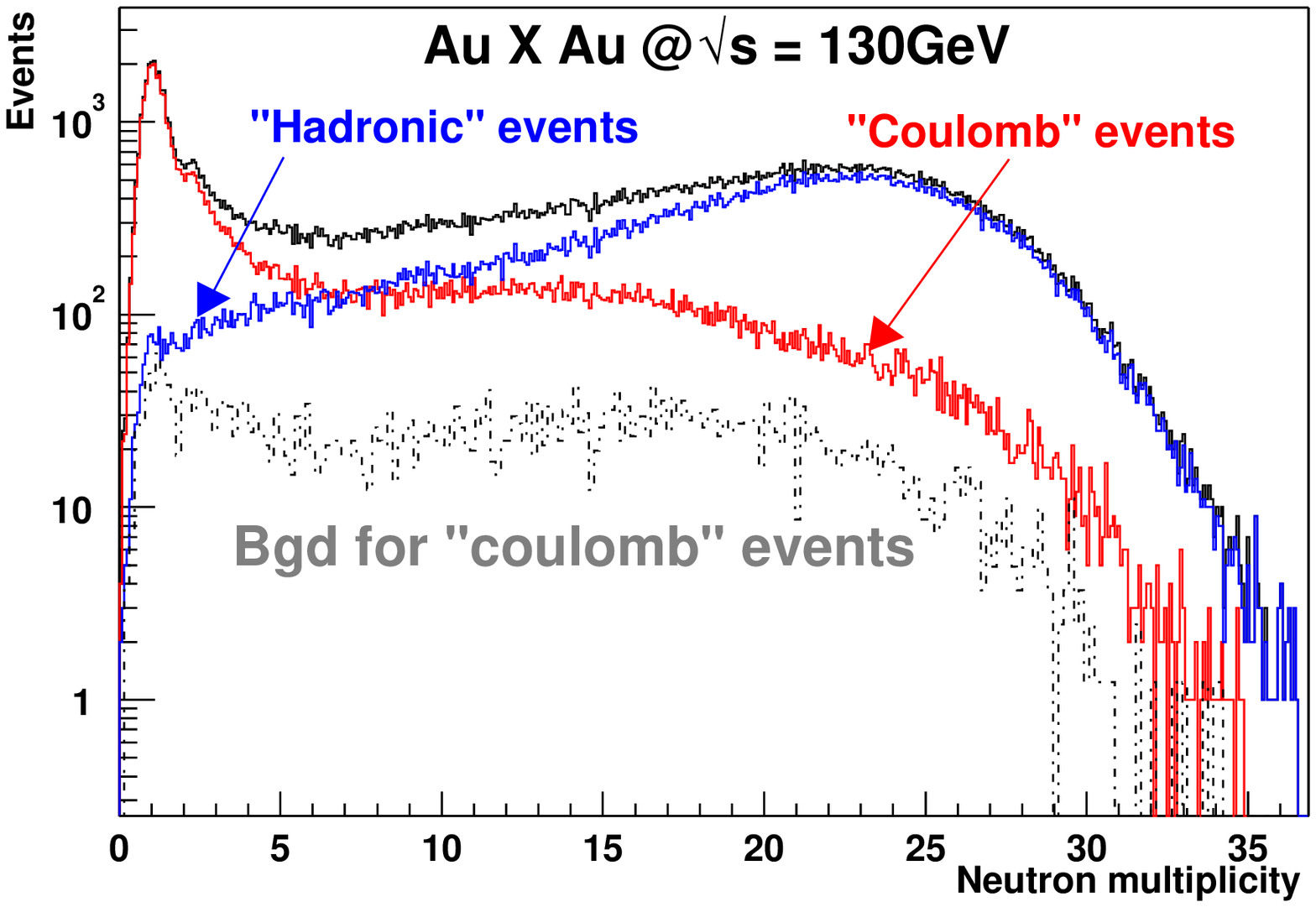}
}
\vspace*{-1cm}
\caption{\small Spectrum of a single ZDC detector (from PHENIX).}
\label{zdc_spectrum}
\end{minipage}
\vspace*{-.75cm}
\end{figure}

To understand whether particle production in nuclear collisions is 
fundamentally different than in $\pp$ or $\pA$ collisions, it is
important to understand the collision geometry.
The impact parameter of the collision determines $\np$, the number of 
nucleons that interact inelastically (i.e. participate), as well
as $\ns$, the number of spectator nucleons which do not interact and 
continue along the beam direction.
To determine $\np$
in fixed target experiments, a zero-degree calorimeter (ZDC) is used
to measure the forward energy and directly infer $\np$ by the relationship
$\np + \ns = 2A$.
At RHIC, each experiment is equipped with a pair of ZDCs\cite{zdc} 
placed behind dipole magnets which sweep away charged particles
and nuclear fragments, leaving primarily spectator neutrons.
Figure \ref{zdc_vs_pdlmean} shows PHOBOS data on
the relationship between charged particle production in $3<|\eta|<4.5$
and the neutral energy measured by the RHIC ZDCs.
This data suggests that above a certain $\np$, the number
of produced particles is in fact monotonically related to $\np$.

Although one cannot directly measure the number of
participants in a given collision,
one can use this monotonic relationship of $\np$ with respect
to $\nch$ or $\et$
to relate a fraction of the cross section to a range in $\np$.
This is done by means of the Glauber
model of nuclear collisions \cite{glauber,kn},
which allows the calculation of $\np$
as well as the number of binary collisions, $\nc$,
experienced by the colliding system as a function of the impact parameter.

Of course, any attempt to measure a fraction of
the total inelastic cross section is necessarily limited by an
experiment's understanding of its trigger efficiency, since the 
systematic error increases for more peripheral events. 
This is complicated
by the fact that the ZDCs are sensitive to mutual Coulomb dissociation\cite{denisov},
as seen in Figure \ref{zdc_spectrum} from PHENIX.
However, theoretical calculations are available and
current comparisons of data and theory \cite{denisov,katzy,videbeck} 
are in good agreement.

Despite the difficulties in doing so, there are real benefits in estimating
$\np$ and $\nc$.
Scaling $\pA$ and $AA$ results by $\np$ allows the comparison of results 
to $\pp$ data.
Also, experimentalists can correct for the effects of 
fluctuations due to physics effects or detector
acceptance which bias the measurement of the actual nuclear geometry.  
This facilitates making reliable comparisons of the RHIC experiments to each other
and to SPS and AGS data.

\section{Charged Particle Multiplicity}

The multiplicity of charged particles produced in heavy ion collisions
arises from a variety of physics processes.
In addition to the expected soft processes seen at lower energies,
hard processes, nuclear shadowing, and hadronic
rescattering all play a role\cite{hijing}.
Each of these has an effect on the number of degrees of freedom 
available to the colliding system (i.e. the entropy).
Although a value of $\nch$ in isolation does not provide insight
into the relative contributions of the various processes, 
we can attempt to disentangle them by
systematically varying the initial conditions of the collision
and comparing the results to $pp$ and $pA$ data as well as 
theoretical models.  For this, it is useful to study
the charged particle multiplicity per
participant pair, $\dndetanp$. 

Theoretical models of particle production in RHIC collisions 
broadly fall into two classes (see \cite{eskola}).
The first class is based on modifying the wounded nucleon model to include
hard processes.
In these, one 
assumes that hard and soft processes scale with binary
collisions and wounded nucleons, respectively:
$\dndeta = A \times \np + B \times \nc$.
Both HIJING \cite{hijing,hijing-cent} and
the eikonal model proposed by Kharzeev and Nardi (KN) \cite{kn}
follow this approach.  
However, while HIJING includes additional physics effects (jet quenching and
nuclear shadowing) which
lead to a linear rise in $\dndetanp$ vs. $\np$, the KN calculation
only uses as input the fraction of hard processes and the PHOBOS result.
This leads to a dependence of $\dndeta$ similar to that
measured by WA98\cite{wa98} and WA97\cite{wa97} which 
is well described by a simple power-law form, $C \np ^\alpha$.

The other class of calculations, 
based on parton saturation, predict a very different dependence on $\np$.
EKRT \cite{ekrt,eskola} found that a geometry-dependent saturation
scale predicts a nearly-constant dependence of $\dndetanp$ 
as a function of $\np$.
Kharzeev and Nardi \cite{kn} also perform a calculation
based on parton saturation, 
including the running of the strong coupling constant with saturation scale,
that finds $\dndeta$ per participant to
scale as $\ln ( Q^2_s / \Lambda^2 )$, where $Q^2_s$ is the impact-parameter
dependent saturation momentum scale.  This result 
agrees well with their eikonal calculation, perhaps fortuitously.

Before comparing these predictions with experimental data, it is
important to make sure that both theory and experiment are discussing
the same quantity.
To convert $\dndy$ to $\dndeta$,
it is not correct simply to apply a global correction factor, e.g. 95\%.
Rather, one must apply the proper
Jacobian, $dy = \beta d\eta$, to the $\dndy$ distribution for each
species of charged particle
to transform it to $\dndeta$.
It is useful to expand this expression,
\begin{equation}
\frac{dN}{d\eta} = \frac{dN}{dy} \sqrt{1 - 
\left( \frac{m}{m_T \cosh y} \right) ^2},
\label{dndetady}
\end{equation}
where we see that the two distributions converge at large $y$ and $m_T$.
Thus, $\dndeta$ near mid-rapidity generally agrees with $\dndy$
at fixed-target experiments at CERN, but significant deviations ($\sim~15~\%$)
arise at collider experiments, even for lighter species like pions. 

\begin{figure}[t]
\begin{minipage}[t]{75mm}
\includegraphics[width=8cm,height=5cm]{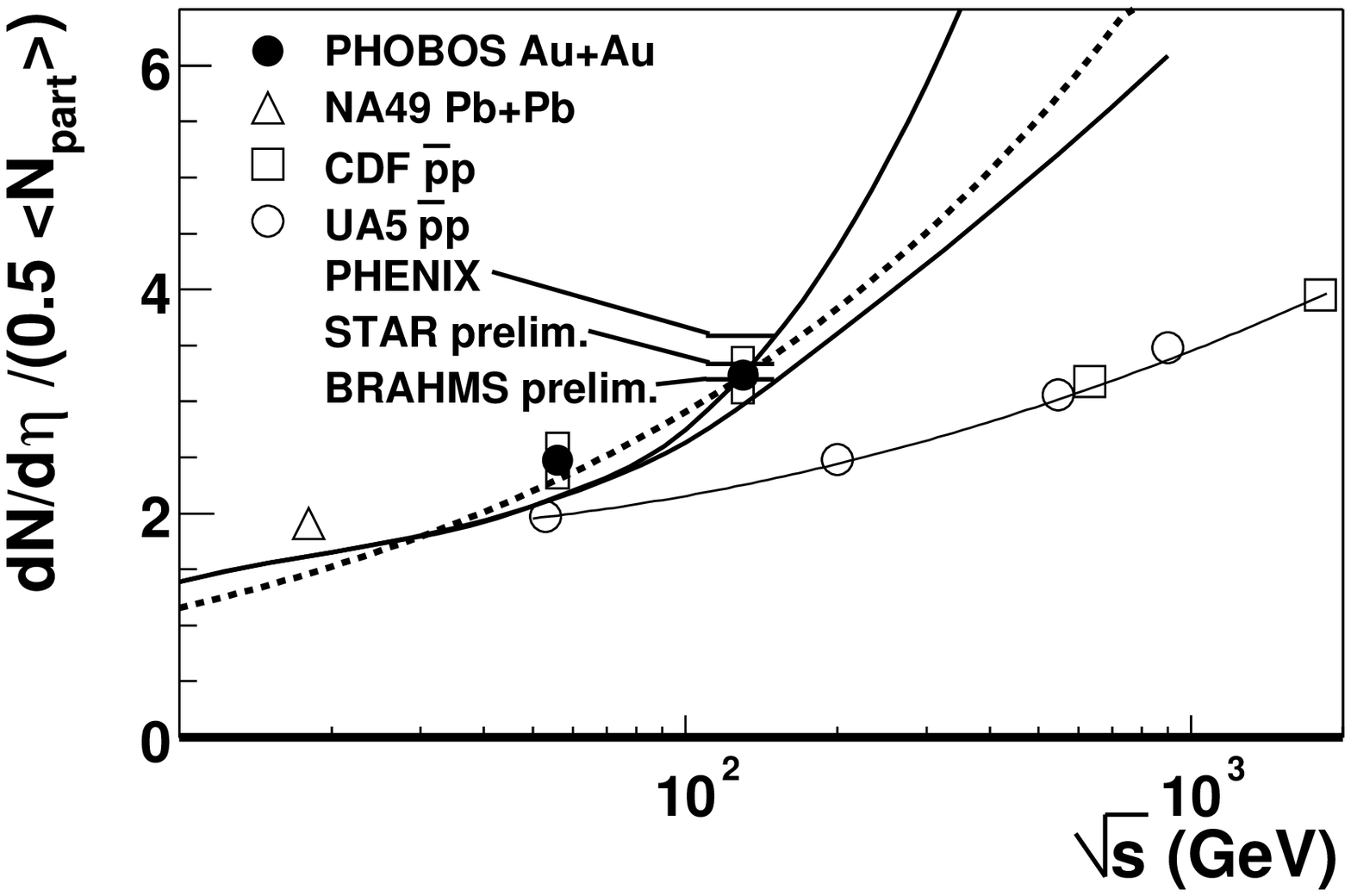}
\vspace*{-1cm}
\caption{\small Energy dependence of $\dndetaonenp$ with results from all four RHIC experiments at $\snn=130$ GeV.
}
\label{dndeta-theory-exp}
\end{minipage}
\hspace{\fill}
\begin{minipage}[t]{75mm}
\includegraphics[width=8cm]{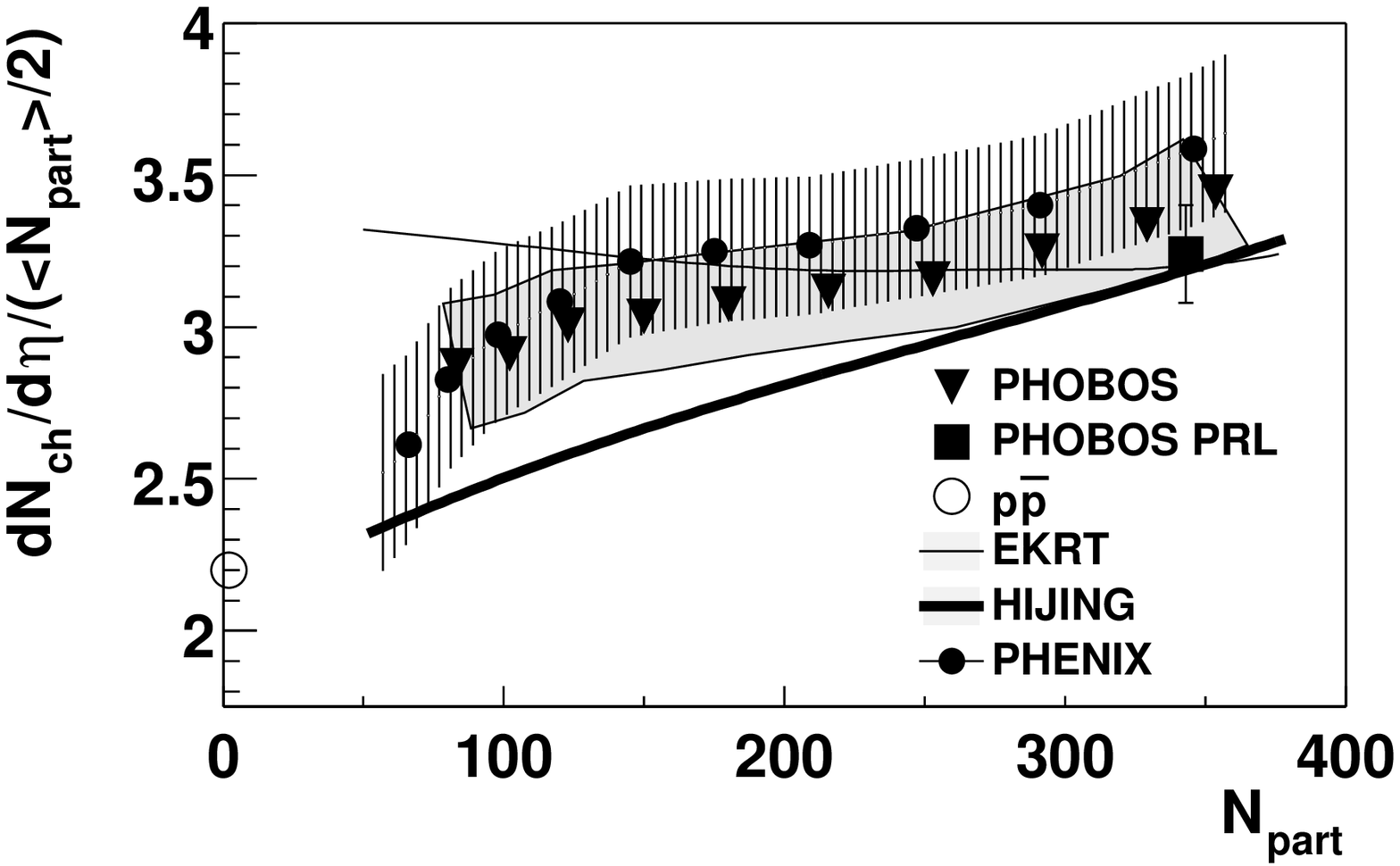}
\vspace*{-1cm}
\caption{\small PHENIX and preliminary PHOBOS results for $\dndetazeronp$}
\label{phobos_phenix}
\end{minipage}
\vspace*{-.5cm}
\end{figure}

The first published results from RHIC were the measurement of $\dndetazeronp$
as a function of energy \cite{phobos-dndeta}, 
shown in Figure \ref{dndeta-theory-exp}.
This provided the first opportunity to compare
extrapolations of $\pp$ and $\pbarp$ collisions to
$AA$ collisions at the highest available energies.
The new results show that 70\% more particles
are produced than at the SPS and 40\% more than the 
extrapolation from $\pbarp$ (also shown) would predict at $\snn =$ 130 GeV.
This is strong evidence that particle production is not simply due to
independent $NN$ interactions.  Instead, whatever process amplifies the 
production at SPS energies relative to $\pp$ collisions 
is even stronger at RHIC.
At Quark Matter, the other three RHIC experiments presented 
new measurements \cite{videbeck,milov,ullrich}
which are consistent with the original PHOBOS value 
within the stated systematic errors.
We also show the predicted energy dependence
from HIJING and an early EKRT calculation (both from \cite{hijing-cent}).
Clearly, the increasing theoretical uncertainties emphasize the importance
of varying the collision energy at RHIC.

At a fixed collision energy, one useful observable
for understanding the relative role of hard
processes is the variation of $\dndeta$ with $\np$.
The results of $\dndetazeronp$ vs. $\np$ is presented in Figure 
\ref{phobos_phenix} which incorporates the results from 
PHENIX\cite{phenix-dndeta} and preliminary data from PHOBOS\cite{katzy}, 
which agree very well within systematic errors.
These measurements cut off at $\np \sim 70$ since the systematic
error on $\np$ grows rapidly below this 
(a situation which could be improved by colliding smaller nuclei, 
where even central events would have a smaller $\np$ than Au+Au).
Interestingly, the data disfavors
both of the leading predictions proposed before Quark Matter.
Rather than being constant or rising linearly with $\np$, as predicted
by EKRT and HIJING respectively, the dependence on $\np$ looks most similar to
that measured at the SPS and thus is in broad 
agreement with the Kharzeev-Nardi calculations\cite{kn}. 
 
\begin{figure}[t]
\begin{minipage}[t]{75mm}
\includegraphics[width=8cm]{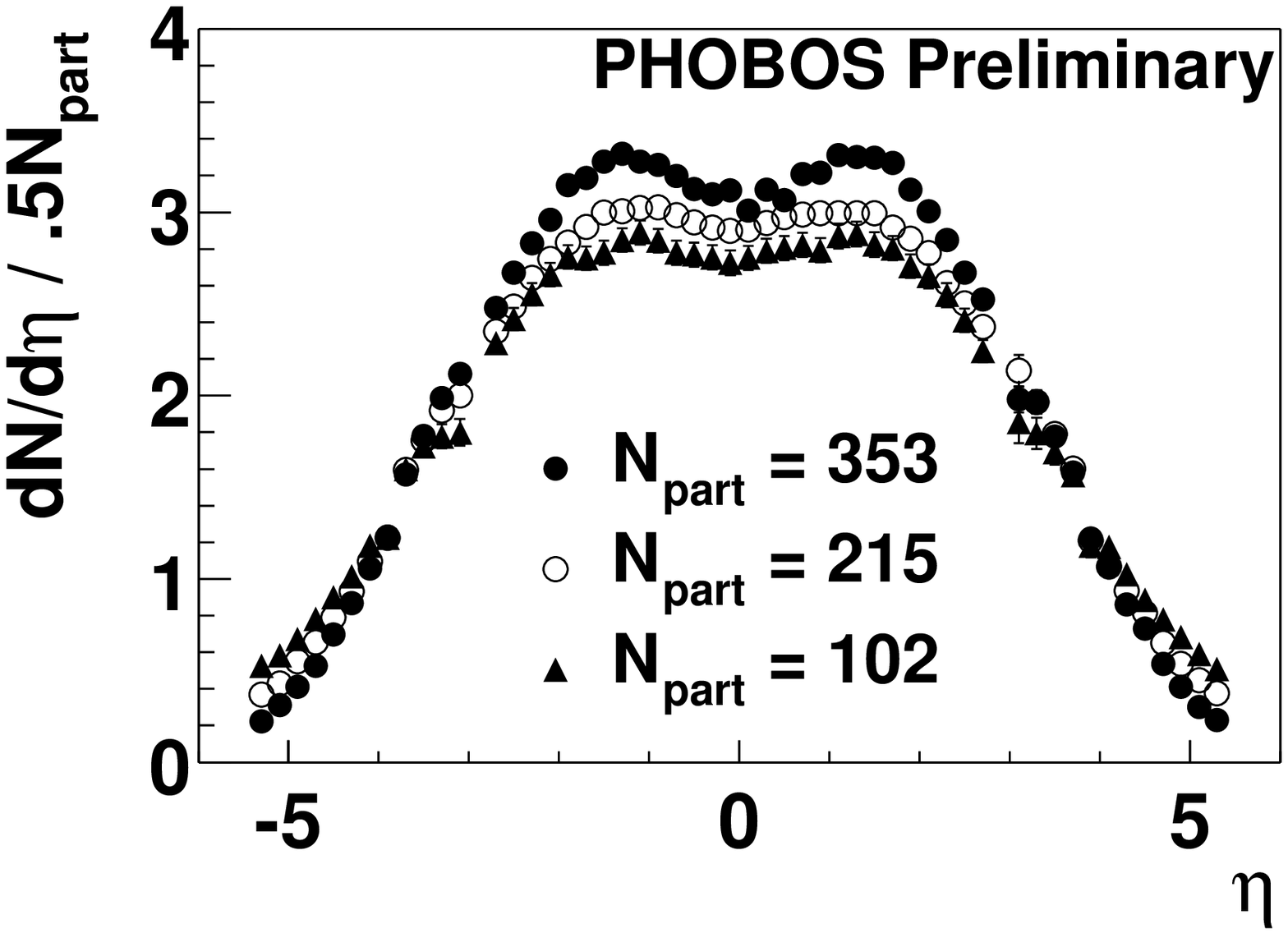}
\vspace*{-1cm}
\caption{\small PHOBOS $\dndeta$ distributions scaled by the number of participants.}
\label{phobos_dndeta_scaled}
\end{minipage}
\hspace{\fill}
\begin{minipage}[t]{75mm}
\includegraphics[width=8cm,height=5cm]{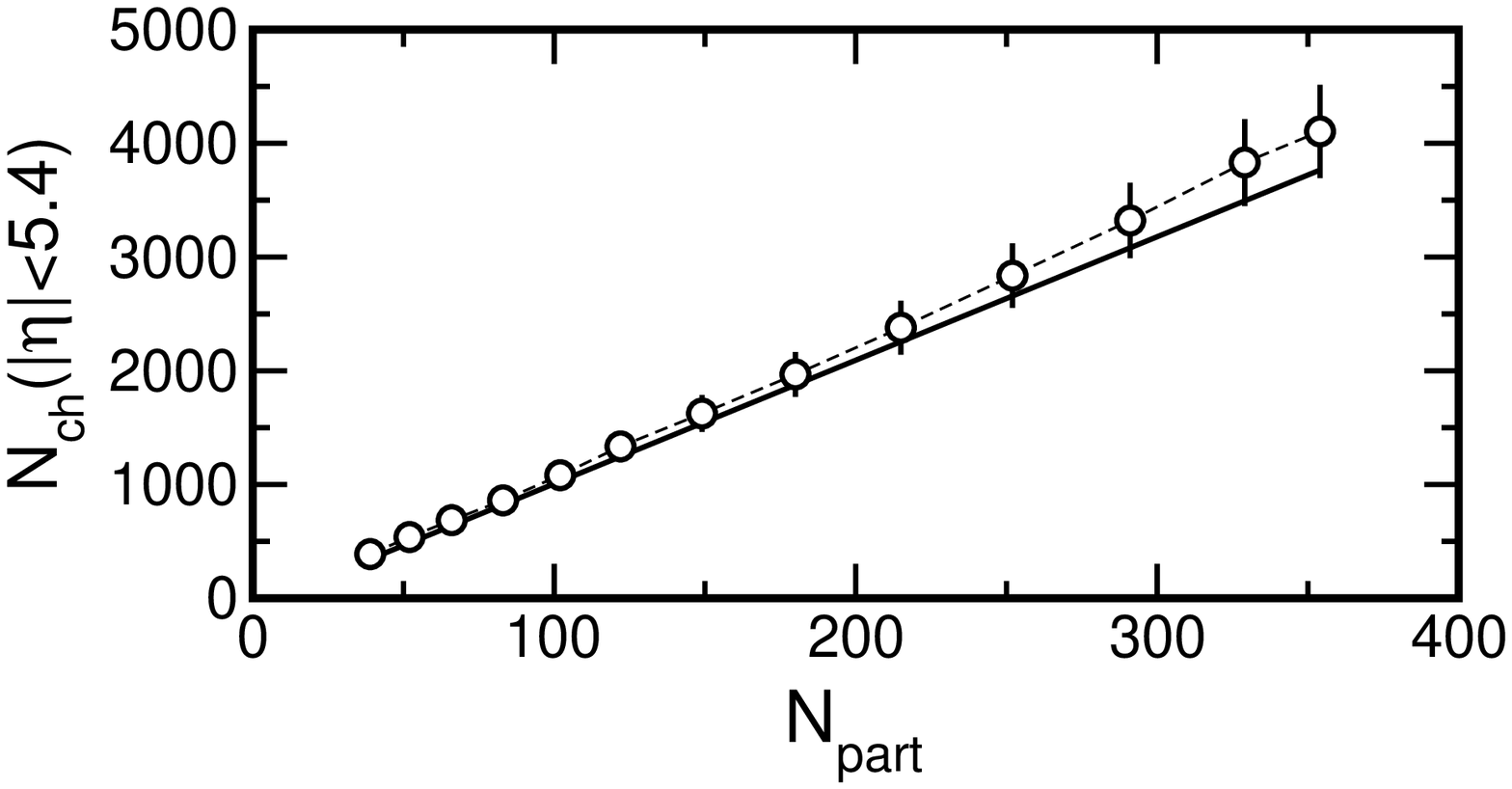}
\vspace*{-1cm}
\caption{\small Preliminary PHOBOS data on total charged particle multiplicity vs. $\np$}
\label{total_vs_npart}
\end{minipage}
\vspace*{-.75cm}
\end{figure}

One can also study the full distribution of $\dndeta$, which is sensitive
to all of the abovementioned physics effects but also gives insight
into the role of longitudinal expansion and 
hadronic rescattering \cite{wuosmaa}.
Preliminary PHOBOS data on $\dndetanp$ scaled by $\halfnp$ is shown
in Figure \ref{phobos_dndeta_scaled}.  
As $\np$ increases, the distribution gets narrower.
Indeed, forward particle production per participant actually {\it decreases},
suggesting that some fraction of particles are pulled back towards
mid-rapidity via rescattering.  Events generated with
RQMD 2.4\cite{rqmd}, which 
contains a substantial amount of such rescattering, qualitatively
verifies this hypothesis.  This seems to contradict models such
as HIJING which have no additional evolution of the system after the
initial particles are produced, leading to no change in the particle
production per participant outside of $|\eta|>3$.

By integrating the $\dndeta$ distributions, PHOBOS also presented
a measurement of the total number of charged particles 
as a function of $\np$\cite{wuosmaa}, shown in Figure \ref{total_vs_npart}.  
Results from a HIJING calculation are also presented and are consistent
within the 10\% systematic errors.
This is remarkable considering the variety of possible physics effects 
that arise out of the collision dynamics.

\section{Transverse Energy}
The measurement of transverse energy ($\et$) gives similar 
information as that of the charged multiplicity measurements, but
$\et$ is also sensitive to the charged particle momentum spectrum.
This gives us access to physics inaccessible to the 
previous measurements:
the mean $\pt$,
the initial energy density (through the Bjorken
formula $\epsilon = d\et/d\eta / \pi R^2 \tau $ \cite{bjorken}),
and the effect of longitudinal flow ($p \smallskip dV$ work) on the evolving system.

\begin{figure}[t]
\begin{minipage}[t]{75mm}
\includegraphics[width=8cm]{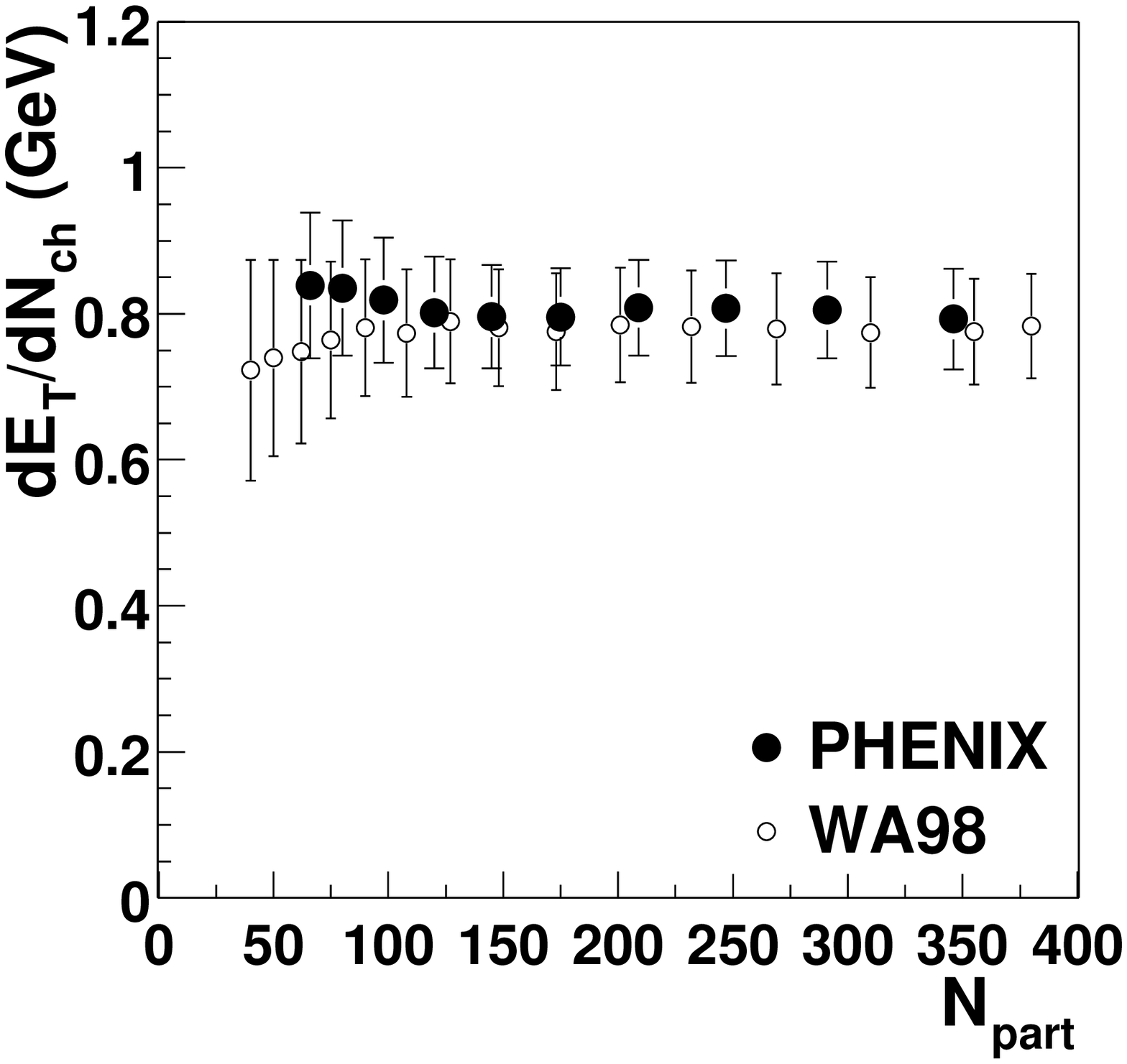}
\vspace*{-1cm}
\caption{\small Preliminary PHENIX data and WA98 data on $\et/\nch$ as a function of $\np$.}
\label{etnch_vs_npart}
\end{minipage}
\hspace{\fill}
\begin{minipage}[t]{75mm}
\includegraphics[width=8cm]{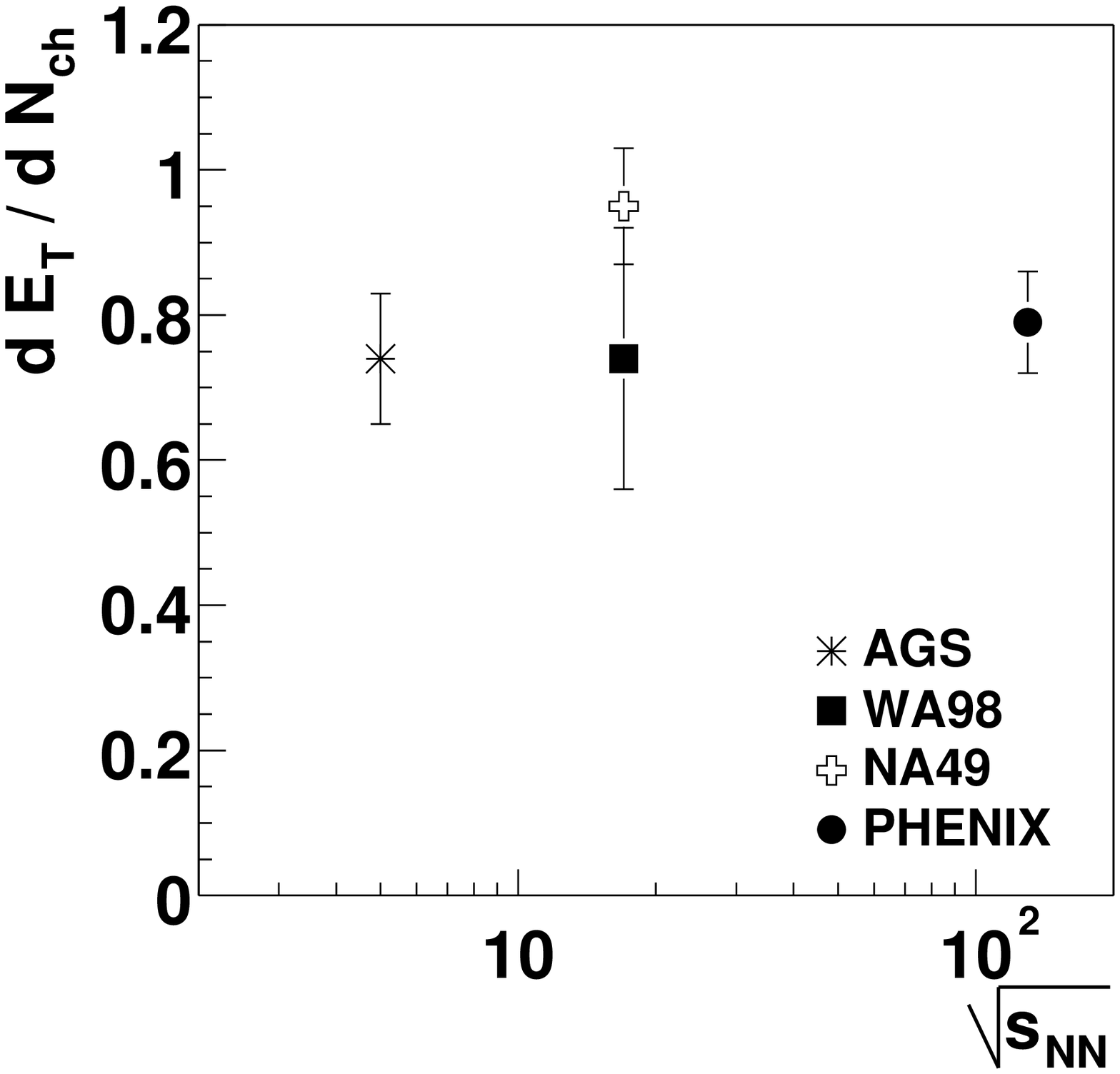}
\vspace*{-1cm}
\caption{\small $\et/\nch$ as a function of $\snn$.  Along with AGS and SPS results, preliminary PHENIX data is shown.}
\label{etnch_vs_energy}
\end{minipage}
\vspace*{-.7cm}
\end{figure}

At QM2001, PHENIX presented three results\cite{milov}
pertaining to the produced transverse energy:
$d\et/d\nch$ vs. $\np$,
$d\et / d\nch$ vs. $\snn$,
and
$d\et / d\eta / \halfnp$ vs. $\np$.
The transverse energy per charged particle is shown in Figure
\ref{etnch_vs_npart}. It appears to be constant
over the full range of centrality, which 
suggests that there is no dramatic modification
of the particle spectra occurs when moving from peripheral to
central collisions.
It also suggests that the transverse
energy at mid-rapidity should scale as the charged particle
multiplicity.  This is confirmed by the measurement of total transverse
energy scaled by $\np$\cite{milov}.
However, while the transverse energy appears to scale simply with the number
of particles,
the transverse energy per charged particle
as a function of $\sqrt{s_{NN}}$, shown in Figure \ref{etnch_vs_energy},
shows some surprising behavior.
It appears that $d\et/d\nch$ is the same at the
AGS and at RHIC.  However, the two experiments at the SPS (WA98 and NA49)
suggest that this number might be the same, or perhaps larger, at SPS energies.  
This would seem to contradict the observations
of the STAR experiment\cite{calderon} 
at this conference that $\langle \pt \rangle$ increases by
$20\%$ from SPS to RHIC energies.  However, other results presented
at the conference, e.g. jet-quenching, imply that particle production
at RHIC shows new features that might lead to deviations from expected
behavior.

One of the most pressing questions at this conference concerns the
energy density achieved in RHIC collisions.  Using the Bjorken formula,
NA49 extracted an initial energy density of
$\epsilon_{BJ} \approx 3.2$ GeV/fm$^3$\cite{na49-et}.  
Applying the same reasoning to
the RHIC experiments would give a $70\%$ increase.  However, while the
measurement of $\et$ seems straightforward, the Bjorken formula has
theoretical uncertainties which do not decrease with beam energy.  
For example, the models of particle production based on parton saturation
have formation times which vary as the inverse of the saturation scale
$\tau \sim \hbar / Q_s \sim .1-.2$ fm.  This would give an energy density
of $16-20 GeV/fm^3$, as shown in \cite{kn}.  Clearly, further 
theoretical investigations
are warranted in order to reduce the ambiguity in the concept of initial
energy density.

\section{Elliptical Flow}
One of the most striking results from the initial round of RHIC results
is the magnitude of the elliptical flow reported by the STAR experiment
\cite{star-flow}, which approaches the levels predicted by hydrodynamic
calculations \cite{olli92}.
Since this initial result, the other experiments have
confirmed this data and new information is now available about other
aspects of this phenomenon\cite{snellings,park,lacey}.

\begin{figure}[tb]
\begin{minipage}[t]{75mm}
\raisebox{-.5cm}
{
\includegraphics[width=7cm]{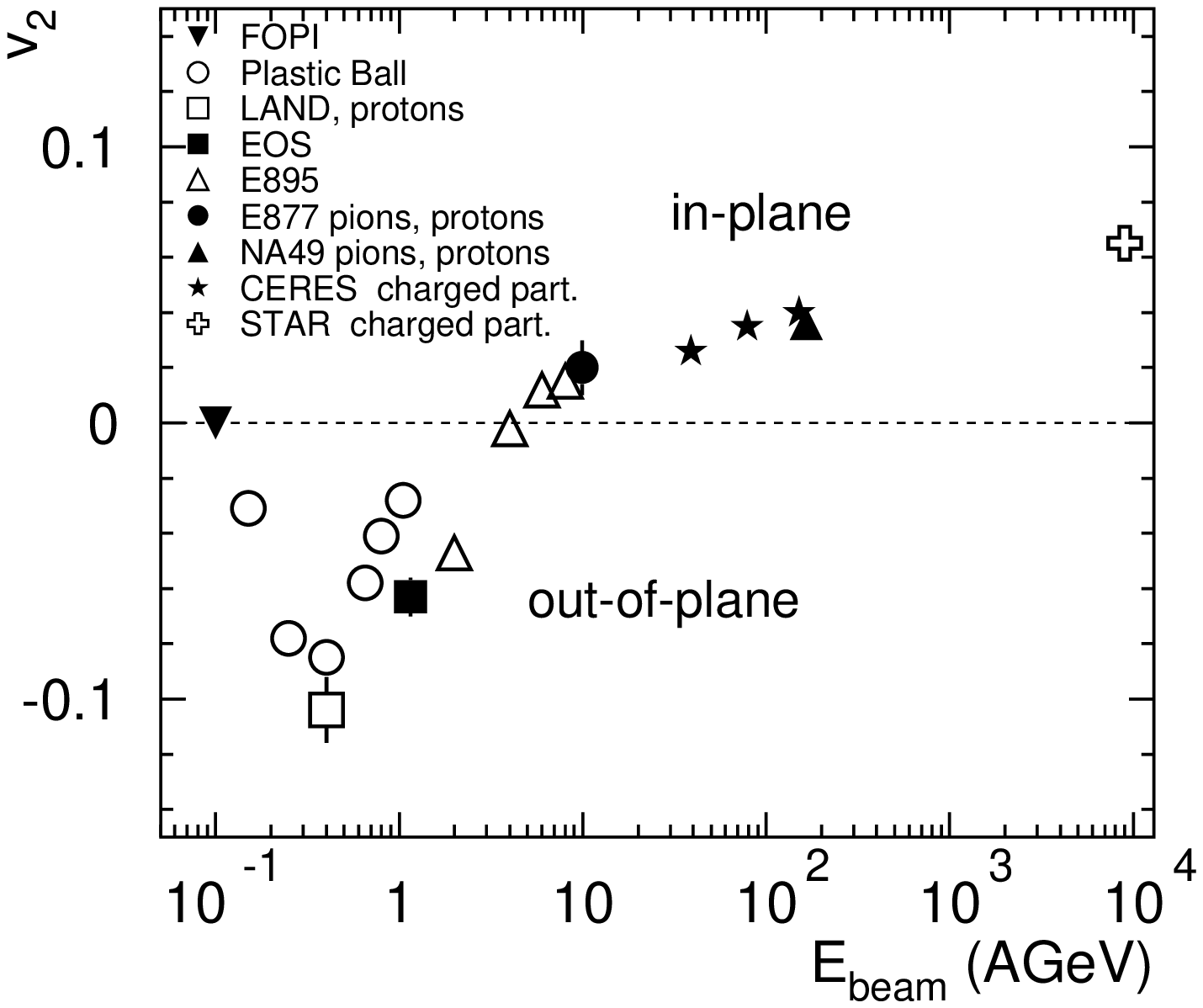}
}
\vspace*{-1cm}
\caption{\small Energy dependence of $\vtwo$, showing data
ranging from Bevalac to RHIC energies.  }
\label{energy_flow}
\end{minipage}
\hspace{\fill}
\begin{minipage}[t]{75mm}
\includegraphics[width = 8cm, height = 6cm]{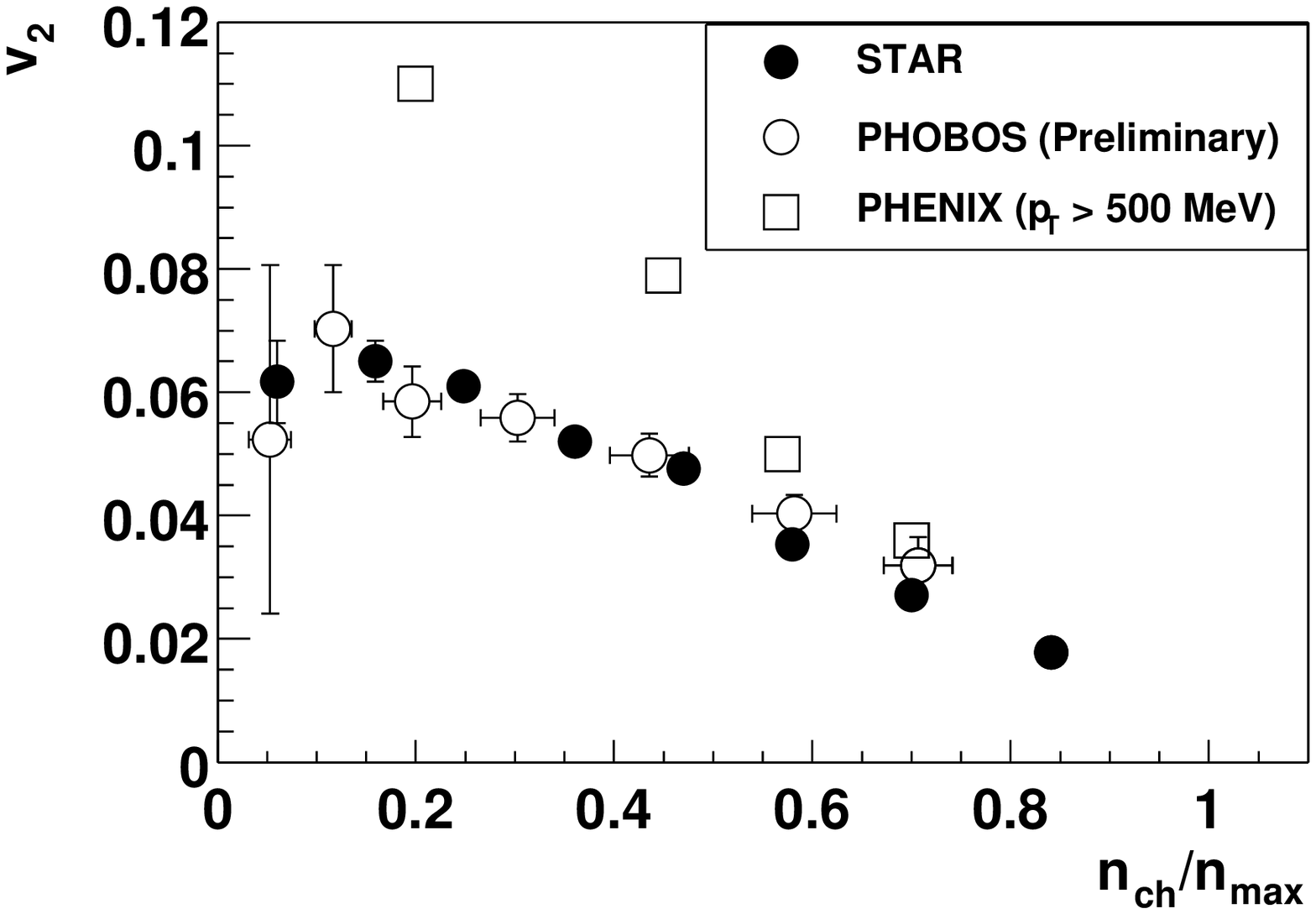}
\vspace*{-1cm}
\caption{\small The STAR result compared with PHOBOS and PHENIX.  PHENIX has a lower $\pt$ cutoff of 500 MeV.}
\label{all_flow}
\end{minipage}
\vspace*{-.7cm}
\end{figure}

As the two nuclei collide, the vector joining their
centers defines the ``reaction plane''.
While one cannot directly measure the true reaction plane, 
subevent analysis is used to determine the experimental
resolution obtained by using the measured particles themselves
to estimate the collision angle $\Psi$,
$n \Psi = \arctan ( \sum_i{\sin(n\phi_i)} / \sum_i{\cos(n\phi_i)} )$
\cite{snellings}.
Once $\Psi$ is determined, one can extract the Fourier components
$\vone$ corresponding to directed flow, and $\vtwo$ corresponding to
the magnitude of elliptic particle flow. 
For a detector with limited acceptance, such as 
the PHENIX experiment with $\Delta \phi = 180^o$,
one can use the two-particle correlation function to extract $\vtwo$,
$\frac{dN}{d\Delta\phi} \propto 1 + \sum^{\infty}_{n=1}{2 v_n^2 \cos(n\Delta\phi)}$
\cite{lacey}.
In principle, these two methods extract 
the same information.  However, the correlation function method
may be sensitive to other effects, e.g. jets or HBT, that
would be missed by the other method.

The magnitude of $\vtwo$ varies with
the energy as well as the centrality of the collision.
The centrality dependence is controlled by the eccentricity $\epsilon$
of the nuclear overlap region.  
It has been shown \cite{vpPLB,heiselberg} that when the system is dilute,
$\vtwo \propto \epsilon \times dN/dy$, where the rapidity density characterizes
the probability of particles to rescatter.
In the limit where the number of scatterings becomes large, 
only the initial geometry is important, so $\vtwo \propto \epsilon$,
and the proportionality constant can be predicted via hydrodynamic
calculations\cite{olli92}.
The compiled data for the energy dependence of $\vtwo$ is shown in Figure
\ref{energy_flow} (from \cite{appelshauser}).   
The large anti-flow (squeeze-out)
observed in low-energy nuclear collisions is seen to change sign at 
$\snn = 4 $GeV, turning into a continuous logarithmic rise of $\vtwo$ 
all the way to RHIC energies.
It is interesting to note that
$v_2$ at RHIC is approximately 60\% higher than at the SPS, 
similar to the 70\% increase in $\dndeta$ already mentioned.

Since STAR's original result \cite{star-flow},
three of the four RHIC experiments have measured the dependence
of $\vtwo$ with centrality, as shown in Figure \ref{all_flow}.  
Both STAR and PHOBOS measure the event
with full azimuthal acceptance and comparable event plane resolution.
PHENIX uses the correlation function 
method and has a lower $\pt$ cutoff of 500 MeV, both of which may
explain why the PHENIX $v_2$ result is somewhat higher than the other two.  

\begin{figure}[tb]
\begin{minipage}[t]{75mm}
\includegraphics[width=8cm]{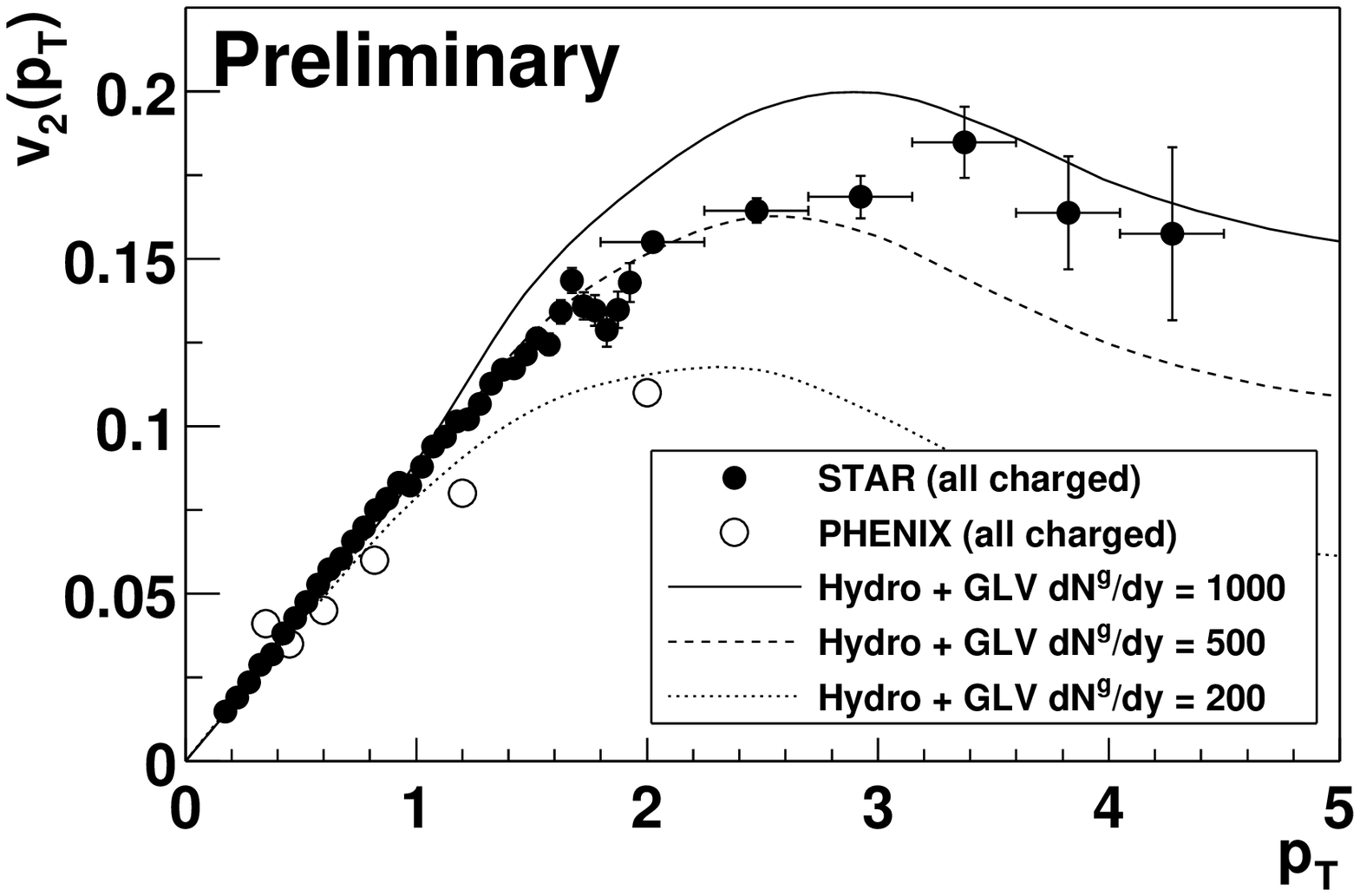}
\vspace*{-1cm}
\caption{\small Comparison of charged particle $\vtwo(\pt)$ with GLV calculations including hydro and jet-quenching.  Preliminary data from PHENIX and STAR are shown.}
\label{flow_quenching}
\end{minipage}
\hspace{\fill}
\begin{minipage}[t]{75mm}
\includegraphics[width=8cm]{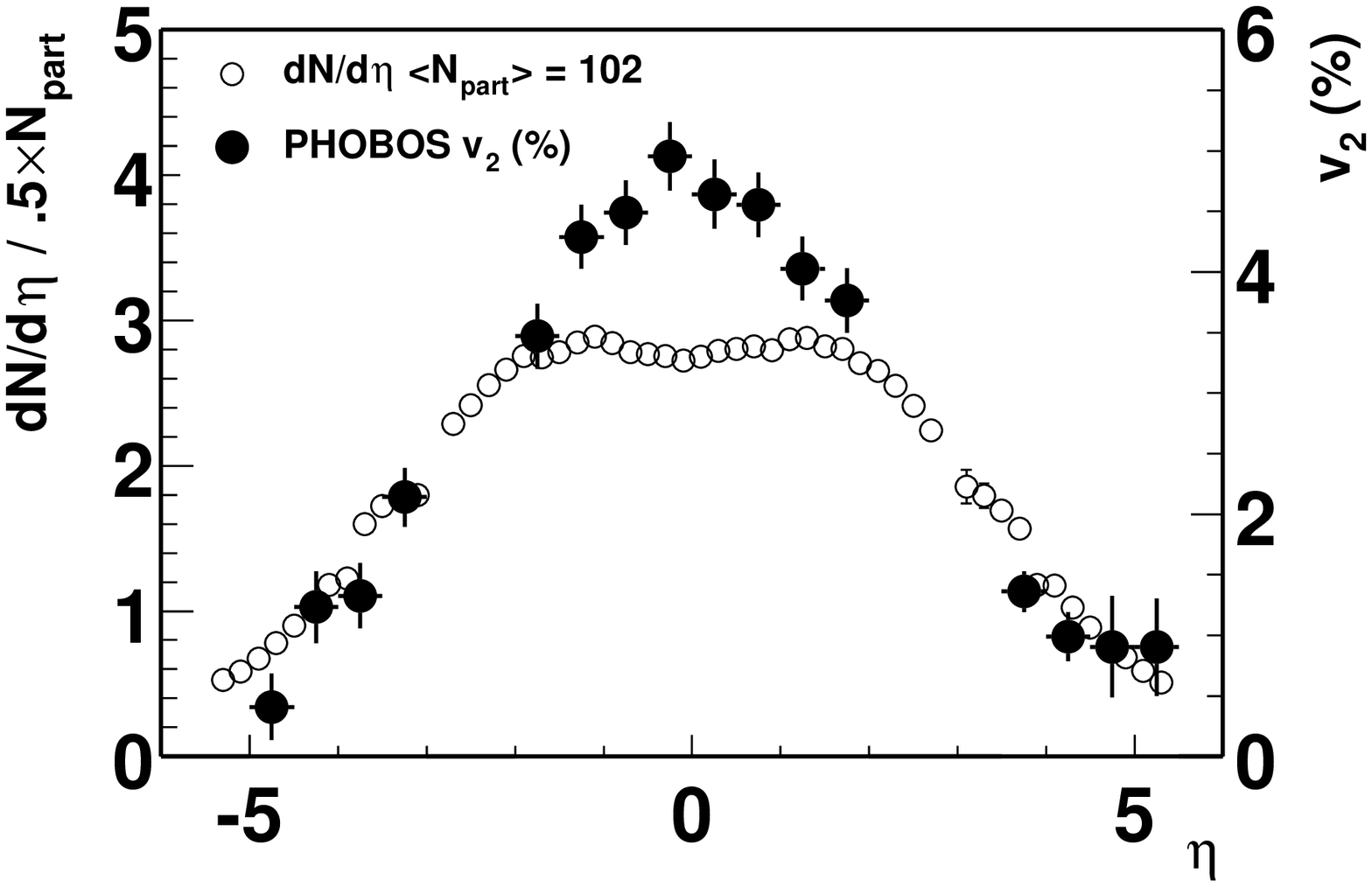}
\vspace*{-1cm}
\caption{\small $\vtwo$ (from PHOBOS) as a function of $\eta$ overlaid on top of $\dndeta$ for $\np = 102$ (see Figure \ref{phobos_dndeta_scaled}).}
\label{flow_eta}
\end{minipage}
\vspace*{-.5cm}
\end{figure}

The $\pt$ dependence of $\vtwo$ appears to both support 
the hypothesis that the central region of RHIC collisions shows hydrodynamic behavior,
as well as suggest the appearance of jet quenching.
STAR and preliminary PHENIX data\cite{lacey} on $\vtwo(\pt)$
are shown in Figure \ref{flow_quenching} and are in broad agreement,
at least for the basic trend.
Calculations given in \cite{huovinen} have reproduced the STAR data up to 
$\pt = 2$ GeV.  
The same calculations also predict $\vtwo$ for identified pions and protons 
and finds that protons only have non-zero flow above $\pt \sim 400$ MeV.
Also shown in Figure \ref{flow_quenching} are 
calculations of jet quenching which incorporate
the geometry of the collision zone\cite{glv}.  They find that 
quenching has a dramatic effect on $\vtwo$ at very high-$\pt$.
Both of these are consistent with the preliminary STAR data\cite{snellings}.

PHOBOS presented data for $\vtwo$ vs. $\eta$ all the way to
 $\eta = 5.4$\cite{park}, shown in Figure \ref{flow_eta}.
Although results from three-dimensional hydrodynamic calculations are
not available, one can still ask if $\vtwo$ away from mid-rapidity
scales with $\dndy$, which is similar to $\dndeta$ in this region.
Figure \ref{flow_eta} shows the PHOBOS data overlaid with
the most peripheral distribution from Figure \ref{phobos_dndeta_scaled}.
Following a suggestion by Manly\cite{manly}, one can scale the multiplicity
distribution to roughly match $\vtwo$ for $|\eta|>2$, to find remarkable
agreement in the shape.
The sharper peak in the flow distribution might simply be due to the 
flattening of the multiplicity distribution near $\eta = 0$ 
suggested by the Jacobian shown in equation \ref{dndetady}.

\section{Conclusions}
We have seen that the global and flow observables measured at RHIC
all show non-trivial collective
behavior which was not predicted by any model.
Indeed, the presence of the strong elliptic flow
and the evolution of $\dndeta$ at the forward and backward rapidities show the
effect of rescattering processes that models such as HIJING do not incorporate.
With the upcoming 200 GeV run at RHIC, there are great opportunities to push our
understanding even further.
However, it should not be forgotten that 
extensive species and energy scans
will be needed to make sure that we have an appropriate understanding of the
basic features of particle production 
before we make detailed interpretations of the high statistics RHIC data.

\section{Acknowledgements}
The author would like to thank the organizers of Quark Matter 2001
for a stimulating experience on all fronts.  
Thanks to the RHIC collaborations and especially
H. Appelshauser,
M. Chiu,
R. Lacey, 
A. Milov,
I. Park,
R. Snellings,
X.-N. Wang,
and 
S. White
for providing results for this article.


\begin{thebibliography}{9}
\bibitem{qm2001} Proceedings of ``Quark Matter 2001'', to appear in Nucl. Phys. A (referred to here as ``these proceedings'').
\bibitem{zdc} C.~Adler, A.~Denisov, E.~Garcia, M.~Murray, H.~Strobele and S.~White, nucl-ex/0008005.
\bibitem{glauber} R.J. Glauber, in Lectures in Theoretical Physics, edited by W.E. Brittin and L.G. Dunham (Interscience, N.Y., 1959), Vol. 1, 315.
\bibitem{kn} D.~Kharzeev and M.~Nardi, nucl-th/0012025.
\bibitem{denisov} A.~Denisov, these proceedings. 
\bibitem{katzy} J.~Katzy, these proceedings.
\bibitem{videbeck} F.~Videbeck, these proceedings.
\bibitem{eskola} K.~Eskola, these proceedings.
\bibitem{hijing} M.Gyulassy and X.N.Wang, Phys.\ Rev.\ {\bf D44} 3501 (1991).
\bibitem{hijing-cent} X.~Wang and M.~Gyulassy, nucl-th/0008014.
\bibitem{pp} F.Abe et al., Phys. Rev.\ {\bf D41}, 2330 (1990).
\bibitem{phobos-dndeta} B.~B.~Back {\it et al.}, Phys.\ Rev.\ Lett.\ {\bf 85}, 3100 (2000).
\bibitem{wa98} M.~M.~Aggarwal {\it et al.}, Eur.\ Phys.\ J.\ C {\bf 18}, 651 (2001).
\bibitem{wa97} WA97 and NA57 Collaborations, F. Antinori et al.,  Preprint CERN-EP-2000-002.
\bibitem{ekrt} K.~J.~Eskola, K.~Kajantie, P.~V.~Ruuskanen and K.~Tuominen, Nucl.\ Phys.\ {\bf B570}, 379 (2000).
\bibitem{milov} A.~Milov, these proceedings.
\bibitem{ullrich} T.~Ullrich, private communication.
\bibitem{phenix-dndeta} K.~Adcox {\it et al.}, nucl-ex/0012008.
\bibitem{wuosmaa} A.~Wuosmaa, these proceedings.
\bibitem{rqmd} H.~Sorge, nucl-th/9905008.
\bibitem{bjorken} J.~D.~Bjorken, Phys.\ Rev.\ D {\bf 27}, 140 (1983).
\bibitem{calderon} M. Calder\'{o}n, these proceedings.
\bibitem{na49-et} S.~Margetis {\it et al.} Phys.\ Rev.\ Lett.\ {\bf 75}, 3814 (1995).
\bibitem{star-flow} K.~H.~Ackermann {\it et al.}, Phys.\ Rev.\ Lett.\ {\bf 86}, 402 (2001).
\bibitem{olli92}  J.-Y.~Ollitrault, Phys. Rev.~D {\bf 46}, 229 (1992).
\bibitem{snellings} R.J.~Snellings, these proceedings.
\bibitem{park} I.C.~Park, these proceedings.
\bibitem{lacey} R.~Lacey, these proceedings.
\bibitem{vpPLB} S.~A.~Voloshin and A.~M.~Poskanzer, Phys.\ Lett.\ B {\bf 474}, 27 (2000).
\bibitem{heiselberg} H. Heiselberg and A. Levy, Phys.\ Rev.\ {\bf C59}, 2716 (1999).
\bibitem{appelshauser} H.~Appelshauser, these proceedings.
\bibitem{huovinen} P.~F.~Kolb, P.~Huovinen, U.~Heinz and H.~Heiselberg, Phys.\ Lett.\ B {\bf 500}, 232 (2001).
\bibitem{glv} M.~Gyulassy, P.~Levai and I.~Vitev, Nucl.\ Phys.\ B {\bf 571}, 197 (2000).
\bibitem{manly} S.~Manly, private communication.
\end{thebibliography}
\end{document}